\documentclass[11pt,a4paper]{article}
\usepackage{jheppub}
\usepackage{pdflscape}
\usepackage{amsmath}
\usepackage{amssymb}
\usepackage{dcolumn}
\usepackage{bm}
\usepackage{color}
\usepackage{epsfig}
\usepackage{amsfonts}
\usepackage{graphicx}
\usepackage{subfigure}
\usepackage{dcolumn}

\newcommand{\be}{\begin{equation}}
\newcommand{\ee}{\end{equation}}
\newcommand{\bea}{\begin{eqnarray}}
\newcommand{\eea}{\end{eqnarray}}

\def\be{\begin{equation}}
\def\ee{\end{equation}}
\def\bea{\begin{eqnarray}}
\def\eea{\end{eqnarray}}

\begin{document}

\title{ Generalized gravitational phase transition in novel 4D Einstein-Gauss-Bonnet gravity}

\author[a]{Daris Samart,}
\author[b,c,d,e]{Phongpichit Channuie}

\affiliation[a] {Department of Physics, Faculty of Science, Khon Kaen University, Khon Kaen, 40002, Thailand}
\affiliation[b] {School of Science, Walailak University, Nakhon Si Thammarat, 80160, Thailand} 
\affiliation[c] {College of Graduate Studies, Walailak University, Nakhon Si Thammarat, 80160, Thailand}
\affiliation[d] {Research Group in Applied, Computational and Theoretical Science (ACTS), \\Walailak University, Nakhon Si Thammarat, 80160, Thailand} 
\affiliation[e] {Thailand Center of Excellence in Physics, Ministry of Higher Education, Science, \\Research and Innovation, Bangkok 10400, Thailand}

\emailAdd{darisa@kku.ac.th}
\emailAdd{channuie@gmail.com}

\abstract{In this work, we present the possible existence of thermal phase transition between AdS to dS asymptotic geometries in vacuum in the context of novel 4D Einstein-Gauss-Bonnet (EGB) gravity. The phase transition proceeds through the thermalon (the Euclidean sector of the bubble thin-shell) formation having a black hole inside the interior of dS spacetime and a thermal AdS spacetime in the exterior without introducing any matter field. According to our analysis, we find that the gravitational phase transitions using novel 4D EGB gravity always take place for all existences of the 4D EGB coupling. In contrast to 5-dimensional EGB gravity, the existence of phase transition requires a narrow range of the 4D EGB coupling with particular critical values, while the 5-dimension sector has a wider range of its coupling for the existences of the thermalon and there also exist critical values for the emergence of the phase transition.}


\maketitle

\section{Introduction}

The positive value of the cosmological constant or,
equivalently, the energy density of the vacuum, has its original roots, both in cosmology and in gravitational physics. A true mechanism describing the positiveness is yet unknown. However, gravitational phase transition between two competing vacua of a given theory with different cosmological constants possesses one of possible explanations. As is well known,  the simultaneous existence of different AdS/dS vacua is a common feature in various
gravitational theories. For instance, the coupling of the metric to scalar fields \cite{Linde:1974at,Veltman:1974au} or $p$-form fields \cite{Aurilia:1980xj,Duff:1980qv} may give rise to a positive and non-zero vacuum expectation values of the corresponding fields which contribute to the vacuum energy density or cosmological constant. In addition, there are other several mechanisms that may proceed phase transitions between distinct vacua. One of them may be driven by a quantum tunneling process through an instanton \cite{Brown:1987dd,Brown:1988kg}, or it is developed from the thermally activated phase transition \cite{Linde:1977mm,Linde:1980tt,Linde:1981zj} possibly through thermalon (the Euclidean sector of the bubble thin-shell) mediated mechanism \cite{Gomberoff:2003zh,Kim:2007ix,Gupt:2013poa}.

Moreover, the existence of several vacua in physics is indeed a well-known feature of higher-curvature theories of gravity \cite{Cvetic:2001bk,Nojiri:2001pm,Camanho:2012da,Camanho:2015zqa} including a Lovelock one \cite{Camanho:2013uda}. Lovelock gravities provide an
salient framework from a theoretical point of view for many reasons. As higher-order dimensions of curvature, they are useful to explore several conceptual issues of gravity, among them, including physics of black holes. The simplest Lovelock gravity is the time-honored Gauss-Bonnet term \cite{Lovelock:1971yv}, that encompasses a non-trivial dynamics for the higher-dimensional ($D>4$) theories of gravity yielding second-order field equations and it has been studied in the
literature in great detail. Nevertheless, in a four-dimensional description ($D=4$), the Gauss-Bonnet term is topological invariant, and hence does not contribute to the gravitational dynamics.

Very recently, the four-dimensional EGB theory was introduced in Ref.\cite{Glavan:2019inb} by rescaling the Gauss-Bonnet coupling constant $\lambda$\footnote{In existing articles, people have frequently used $\alpha$ instead of $\lambda$.} as $\lambda/(D-4)$, and taking
the limit $D\rightarrow 4$ at the level of the field equation.\footnote{It was noticed that the solution presented by Ref.\cite{Glavan:2019inb} actually was also found in previous papers \cite{Cai:2009ua,Cai:2014jea} for gravity with a conformal anomaly.} Then, the resulting EGB term embodies non-trivial dynamics
in four dimensions. Since then, the outbreak of controversy begins in terms of thorough examinations of the proposed scenario against its validation. In Ref.\cite{Aragon:2020qdc}, the authors study the perturbative and non-perturbative in the coupling constant quasinormal modes of the 4D EGB black hole. Apparently, it has been explored to a large extent. On the one hand, testing the nature of novel 4D Gauss-Bonnet gravity by non-rotating \& rotating, charged \& uncharged black holes gains a major attention \cite{Konoplya:2020bxa,Fernandes:2020rpa,Wei:2020ght,Kumar:2020owy,Konoplya:2020qqh,Ghosh:2020vpc,Singh:2020xju,HosseiniMansoori:2020yfj,Konoplya:2020juj,Kumar:2020uyz,Ghosh:2020syx,Wei:2020poh,Kumar:2020xvu,Zhang:2020sjh,Yang:2020jno}. Additional, we found that it was scrutinized by different approaches. These include geodesics motion and shadow
\cite{Konoplya:2020bxa}, gravitational lensing \cite{Islam:2020xmy,Jin:2020emq}, relativistic stars in 4D EGB \cite{Doneva:2020ped}, and even wormhole constructions \cite{Liu:2020yhu,Jusufi:2020yus}. It is first found in Ref.\cite{Guo:2020zmf} that a negative GB coupling constant is allowed to retain a black hole in 4D EGB gravity, see also Ref.\cite{Zhang:2020qam} for the study of the greybody factor and power spectra of the Hawking radiation. The thin accretion disk around a 4D EGB black hole is studied in Ref.\cite{Liu:2020vkh}. Moreover, the study the bulk causal structure of 4D Einstein-Gauss-Bonnet-Maxwell in the AdS space is found in Ref.\cite{Ge:2020tid}. On the other hand, however, particular arguments on the novelty of the 4D EGB theory was raised in Ref.\cite{Gurses:2020ofy,Hennigar:2020lsl,Bonifacio:2020vbk}. Using the dimensional-regularization approach, the authors of Ref.\cite{Fernandes:2020nbq} claimed that it is not a complete theory, see also Ref.\cite{Mahapatra:2020rds}. An inconsistence of the novel 4D Einstein-Gauss-Bonnet gravity by considering a quantum tunneling process of vacua is claimed by Ref.\cite{Shu:2020cjw}.

The main purpose of this work is to study the possible existence of thermal phase transition between AdS to dS asymptotic geometries in vacuum in the context of novel 4D Einstein-Gauss-Bonnet (EGB) gravity. It was found so far that a study of gravitational AdS to dS phase transition mediated by thermalon has been done in various circumstances \cite{Camanho:2012da,Camanho:2015zqa,Camanho:2013uda,Camanho:2015ysa,Hennigar:2015mco,Sierra-Garcia:2017rni}. However, these existing references were considered in the vacuum solution of the Einstein-Gauss-Bonnet gravity. As mentioned in Ref.\cite{Camanho:2015zqa}, the phase transition of the asymptotically AdS to dS geometries is a generic behavior emerging in the higher order gravitational theories without invoking any matter fields. Interestingly, the phase transition in the presence of the Maxwell field with the fixed charged ensemble is studied in Ref.\cite{Samart:2020qya}. The results behave like the impurity substitution in condensed matter physics.

The structure of the present paper is organized as follows. In section \ref{model}, we take a short survey of providing the full gravitational action of EGB gravity in $d$-dimensions including its boundary terms that are invoked to study the phase transition. Here we also construct a spherically symmetric bubble thin-shell or the thremalon for analysing its dynamics and stability. Particularly in this section, we take a rescaling of the EGB coupling in order to obtain its 4D description. The possible existences of the thermalon mediated phase transition and the relevant thermodynamic quantities in novel 4D EGB gravity from asymptotically AdS to dS asymptotics are given in detail in section \ref{sec3}. We conclude our findings in the last section.

%%%%%%%%%%%%%%%%%%%%%%%%%%%%%%%%
\section{Setup \label{model}}
%%%%%%%%%%%%%%%%%%%%%%%%%%%%%%%%
\subsection{Novel 4D Einstein-Gauss-Bonnet gravity  \label{4DEGB}}
We begin with the full gravitational action of EGB gravity in $d-$dimensions including its boundary terms that are invoked to study the phase transition. The action takes the form
\begin{eqnarray}
\mathcal{I} &=& \int d^dx\left[ -\,\varepsilon_\Lambda\,\frac{(d-1)(d-2)}{L^2} + R + \lambda\,L^2\,\Big( R^2 - 4\,R_{ab}\,R^{ab} + R_{abcd}\,R^{abcd} \Big) \right] 
\nonumber\\
&& -\, \int_{\mathbb{\partial M}} d^{(d-1)}x\sqrt{-h}\left[ K + 2\,\lambda\,L^2
\left\{ J - 2\left( \mathcal{R}^{AB} -\frac12\,h^{AB}\,\mathcal{R}\right)K_{AB}\right\}\right] \,,
\label{5EGBM-action}
\end{eqnarray}
where $L$ is (A)dS curvature radius and here $L^2$ is absorbed to the Gauss-Bonnet coupling $\lambda$ giving the $\lambda$ as the dimensionless parameter. In addition, $J\equiv h^{AB}\,J_{AB}$ is the trace of $J_{AB}$ built from the extrinsic curvature of the hypersurface, $K_{AB}$ via 
\begin{eqnarray}
J_{AB} = \frac13\left( 2\,K\,K_{AC}\,K_B^C + K_{CD}\,K^{CD}\,K_{AB} - 2\,K_{AC}\,K^{CD}\,K_{DB} - K^2\,K_{AB}\right) ,
\end{eqnarray}
and $R_{ab}$ and $\mathcal{R}_{AB}$ are the intrinsic Ricci tensors in the bulk the hypersurface, respectively. Moreover, we have used the indices of the bulk ($d-$dimension) and hypersurface ($(d-1)-$dimension) denoting by small and captital Latin alphabets, respectively, i.e., $a,\,b,\,c,\,\cdots = 0,\,1,\,2,\,\cdots\,,d$ and $A,\,B,\,C,\,\cdots = 0,\,1,\,2,\,\cdots\,,d-1$\,.  
In the following, we make use of the following replacement of the Gauss-Bonnet coupling,
\begin{eqnarray}
\lambda \rightarrow \frac{1}{d-4}\,\lambda_{4}\,.
\end{eqnarray}
This leads to the novel 4D EGB gravity recently proposed in Ref.\cite{Cognola:2013fva,Glavan:2019inb} and in the rest, we will study the gravitational AdS to dS phase transition in the limit $d \to 4$. It has been shown in the literature \cite{Fernandes:2020rpa,Konoplya:2020qqh} that the spherical symmetric solution in $d-$dimension takes the following form,
\begin{eqnarray}
ds^2 = -\,f(r)\,dt^2 + \frac{dr^2}{f(r)} + r^2\,d\Omega^2_{(d-2)}\,,
\end{eqnarray}
where $d\Omega^2_{(\sigma)\,d-2}$ is the line element of the $(d-2)$-dimensional surface of the spatially spherical constant curvature. In the $d\to 4$ limit, the $f(r)$ metric function is given by \cite{Fernandes:2020rpa,Konoplya:2020qqh},
\begin{eqnarray}
f_\pm(r) = 1 + \frac{r^2}{2\,\lambda_{4}\,L^2}\left( 1 \pm \sqrt{1+ 4\,\lambda\left[ 1
+ L^2\frac{\mathcal{M}_\pm}{r^{3}}\right]}\,\right).
\label{metric-f-pm}
\end{eqnarray}
According to the solutions in Eq.(\ref{metric-f-pm}), there are two branches of the solution with different asymptotic geometries. To see these, one considers 
the asymptotic regions by taking $r \to \infty$ and this gives
\begin{eqnarray}
f_\pm(r) &=& 1 - \Lambda_\pm^{\rm eff}\,r^2\,,
\nonumber\\
\Lambda_\pm^{\rm eff} &=& -\left(\frac{1 \pm \sqrt{1+ 4\,\lambda}}{2\,\lambda\,L^2}\,\right).
\label{eff-CC}
\end{eqnarray}
We close this section by discussing the physical implications from the metric $f(r)$ function in Eqs.(\ref{metric-f-pm}) and (\ref{eff-CC}). The minus branch $f_-(r)$ provides the physical solution of the black hole solution and recovers the Einstein gravity for $\lambda\to 0$. While the plus branch $f_+(r)$ solution reveals the naked singulairty and suffers from the Boulware-Deser (BD) ghost instability. In addition, the effective cosmological constant of the plus branch is an unphysical branch because it is infinifty in the limit when $\lambda\to 0$. The effective cosmological constants of the plus and minus branches in Eq.(\ref{eff-CC}) explicitly exhibit the negative and positive values, respectively. This means that the $f_+(r)$ and $f_-(r)$ solutions correspond to AdS and dS spaces, respectively. In the following, we will study the gravitational phase transition mediated by the thermalon between two branches of the solutions and it consequently changes the effective cosmological constant as $\Lambda_+^{\rm eff} \to \Lambda_-^{\rm eff}$.

\subsection{Thermalon dynamics and its stability}
Next we construct a spherically symmetric bubble thin-shell or the thremalon for analysing its dynamics and stability. We start with combining the two different manifolds of the AdS and dS and connect them using the junction timelike hypersurface $\Sigma$ of those two manifolds. The total manifold is composed of $\mathbb{M} = \mathbb{M}_{-} \,\cup\, (\Sigma\times\xi)\, \cup\, \mathbb{M}_+$ where $\xi \in [0,1]$ is the interpolating parameter of two spacetime manifolds. The exterior $\mathbb{M}_+$ and the interior $\mathbb{M}_-$ manifolds correspond to AdS and dS spacetimes, respectively. At $d\to 4$ limit, these two different line elements of AdS exterior manifold ($\mathbb{M}_+$) and dS interior ($\mathbb{M}_-$) are given by
\begin{eqnarray}
ds_\pm^2 = -f_\pm(r_\pm)\,dt_\pm^2 + \frac{dr_\pm^2}{f_\pm(r_\pm)} + r^2_\pm\,d\Omega_{2}^2\,,
\label{out-in-line}
\end{eqnarray}
where the metric function, $f_\pm(r)$ are given by Eq.(\ref{metric-f-pm}). We step further to construct the bubble thin-shell (thermalon) by matching $\mathbb{M}_\pm$ at their boundaries. The boundaries of the manifolds $\partial \mathbb{M}_\pm$ are given by
\begin{eqnarray}
\partial \mathbb{M}_\pm := \Big\{ r_\pm = a \,|\, f_\pm > 0\,\Big\}\,.
\end{eqnarray}
One can parameterize the coordinates as
\begin{eqnarray}
r_\pm = a(\tau)\,,\qquad\qquad t_\pm = \widetilde{t}_\pm(\tau)\,,
\end{eqnarray}
where $\tau$ is comoving time of the induced line elements of the hypersurface which connects the boundaries of both two manifolds $\mathbb{M}_\pm$. The hypersurface line element is written by,
\begin{eqnarray}
ds_\Sigma^2 = -d\tau^2 + a^2(\tau)\,d\Omega_{2}^2\,.
\label{surface-line}
\end{eqnarray}
Matching the line elements of Eqs. (\ref{out-in-line}) with (\ref{surface-line}), we find the constraint
\begin{eqnarray}
1 = f_\pm(a)\left(\frac{\partial\, \widetilde{t}_\pm}{\partial \tau}\right)^2 - \frac{1}{f_\pm(a)}\left(\frac{\partial a}{\partial \tau}\right)^2\,.
\end{eqnarray}
The junction condition of the continuity between two manifolds crossing the hypersurface can be represented in terms of the continuity of the canonical momenta, $\pi_{AB}^\pm$ with the following equation \cite{Camanho:2013uda,Camanho:2015zqa}
\begin{eqnarray}
\pi_{AB}^+ - \pi_{AB}^- = 0\,,
\label{junction-eqn}
\end{eqnarray}
where the canonical momentum, $\pi_{AB}$ is derived by varying the gravitational action of the boundary with respect to the induced metric, $h_{ab}$ on the hypersurface, $\Sigma$, i.e. 
\begin{eqnarray}
\delta \mathcal{I}_{\partial\mathbb{M}} = -\int_{\partial\mathbb{M}}d^{4}x\,\pi_{AB}\,\delta h^{AB}\,.
\end{eqnarray}
It has been shown that all of the canonical momentum components are related to the co-moving time part in the diagonal metric case \cite{Camanho:2013uda,Camanho:2015zqa}. In addition, the co-moving time component $\pi_{\tau\tau}^\pm$ can be expressed in a compact form
\begin{eqnarray}
\Pi^{\pm} = \pi_{\tau\tau}^\pm = \int_{\sqrt{H-g_-}}^{\sqrt{H-g_+}} dx\,\Upsilon'\big[ H- x^2\big]\,,
\label{Pi-time-component}
\end{eqnarray}
where 
\begin{eqnarray}
\Upsilon'[x] = \frac{d\Upsilon[x]}{dx}\,, \qquad \Upsilon[x] = -1/L^2 + x + \lambda_{4} L^2 x^2 \,,\qquad g_\pm(a) = \frac{1 - f_\pm(a)}{a^2},
\end{eqnarray}
and $H = (1 + \dot a^2)/a^2$\,. For convenience, we define a new variable, $\widetilde{\Pi}$, and it is given by $\widetilde{\Pi} = \Pi^+ - \Pi^-$\,. Therefore, the junction conditions of the continuity across hypersurface are fulfilled by the conditions
\begin{eqnarray}
\widetilde{\Pi} = 0 = \frac{d\widetilde{\Pi}}{d\tau}\,,
\end{eqnarray}
These mean that the canonical momenta and their derivatives are continuous at the junction. From then on, we will work on the Euclidean sector transforming $t\,\to\,i\,t$ for studying the thermalon. This provides $\dot a^2 \,\to\,-\dot a^2$ and $\ddot a \,\to\,-\ddot a$\,. The junction condition in Eq. (\ref{junction-eqn}) implies 
\begin{eqnarray}
\widetilde{\Pi} = \Pi_+ - \Pi_- = 0\,\quad \Longrightarrow \,\quad \Pi_+^2 = \Pi_-^2\,.
\label{junction-EGB}
\end{eqnarray}
Using the results of the $\Pi_\pm$ in Eq. (\ref{Pi-time-component}) and the metric tensor $f_\pm(a)$ in Eq.(\ref{metric-f-pm}) in the junction condition above, we can rewrite the junction condition in terms of kinetic energy and effective potential as
\begin{eqnarray}
\Pi_+^2 = \Pi_-^2\,\quad \Longleftrightarrow \,\quad\, \frac12\,\dot a^2 + V(a) = 0\,.
\label{junction-Veff}
\end{eqnarray}
Then the effective potential $V(a)$ of the junction condition equation and its derivative in the $d\to 4$ limit are given by
\begin{eqnarray}
V(a) &=& \frac{a^{5}}{ 24\, \lambda_{4}\,  L^2 \left(\mathcal{M}_+ -\mathcal{M}_-\right)}\left[ \left(1+ 4\,\lambda_{4}\right) g + 4 \left(2 + g\, \lambda_{4}\,  L^2\right)\frac{\mathcal{M}}{a^{3}} \right]\Bigg|_-^+ + \frac{1}{2}\,,
\label{V-eff}
\\
V'(a) &=& \frac{a^4}{24\, \lambda_{4}\,  L^2 \left(\mathcal{M}_+ -\mathcal{M}_- \right)}
\left[5\,(1+ 4\,\lambda_{4})\,g +\left(13  + 2\,g\,\lambda_{4}\, L^2 \right)\frac{\mathcal{M}}{a^{3}}  \right]\Bigg|_-^+ \,,
\label{Vp-eff}
\end{eqnarray}
where we have defined the symbol $\big[ \mathcal{O}\big]\big|_-^+$ as
\begin{eqnarray}
\big[ \mathcal{O}\big]\big|_-^+ \equiv \mathcal{O}_+ - \mathcal{O}_-\,.
\end{eqnarray}
%%%%%%%%%%%%%%%%%%%%%%%%%%%%%%%%%%%%%%%%%%%%%%%%%%%%%%%%%%%%%%%%%%%%%%%%%%%%
\begin{figure}[!h]	
	\includegraphics[width=15cm]{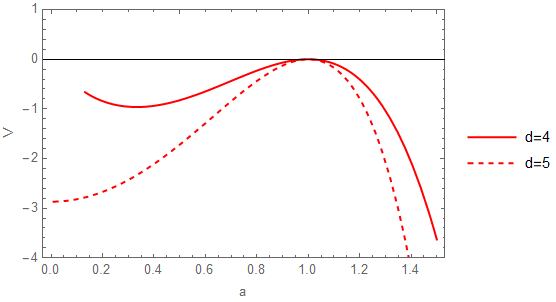}
	\centering
	\caption{The figure displays the shapes of the effective potential of thermalon with $\lambda = 0.05$, $a_\star=1$ and $L=1$. The solid and dashed lines are plots of the effective potetntial for $d=4$ and $d=5$, respectively. }
	\label{thermalon-potential}
\end{figure}
%%%%%%%%%%%%%%%%%%%%%%%%%%%%%%%%%%%%%%%%%%%%%%%%%%%%%%%%%%%%%%%%%%%%%%%%%%%%

%%%%%%%%%%%%%%%%%%%%%%%%%%%%%%%%%%%%%%%%%%%%%%%%%%%%%%%%%%%%%%%%%%%%%%%%%%%%
\begin{figure}[!h]	
	\includegraphics[width=15cm]{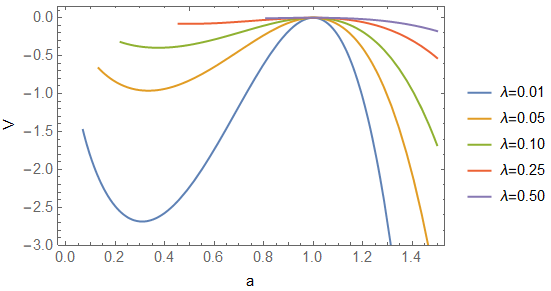}
	\centering
	\caption{The figure displays the shapes of the effective potential of thermalon in several values of the 4D EGB coupling $\lambda_{4}$ with $a_\star=1$ and $L=1$. We note that the effective potential at the location of the thermalon is always unstable in the range of the existent values of the $\lambda_{4}$ coupling and leading to the phase transition always possible.}
	\label{thermalon-potential}
\end{figure}
%%%%%%%%%%%%%%%%%%%%%%%%%%%%%%%%%%%%%%%%%%%%%%%%%%%%%%%%%%%%%%%%%%%%%%%%%%%%
We refer all technical detail derivations of the thermalon effective potential and its derivative in Refs. \cite{Camanho:2013uda,Camanho:2015ysa} for neutral case in general $d-$dimension and Refs. \cite{Samart:2020qya,Samart:2020mnn} for the charged case in 5-dimension. In order to see the shape of the potential with the thermalon configuration, we need to solve the static limit of the thermalon at its location, $a_\star$ via
\begin{eqnarray}
V(a_\star) = 0 = V'(a_\star)\,.
\end{eqnarray}
Having used above conditions, the mass parameters $\mathcal{M}_\pm$ are solved in terms of of $g_\pm^\star$, $a_\star$, $\lambda_4$ and $L$ in $d\to 4$ limit as,
\begin{eqnarray}
\mathcal{M}_+^\star 
&=& \frac{a_\star^{5}\,(1 + 4\,\lambda_4)}{4\, \lambda_4\,L^2\left(3\,a_\star^2 -2\,\lambda_4\,L^2\right)}
\Big[ 4  \left(3 + 2\,\lambda_4\,L^2\,g_-^\star \right) + 20\,\lambda_4\,L^2 \Big]\,,
\label{M_+}\\
\mathcal{M}_-^\star &=& \frac{a_\star^{5}\,(1 + 4\,\lambda_4)}{4\, \lambda_4\,L^2\left(3\,a_\star^2 -2\,\lambda_4\,L^2\right)}
\Big[ 4 \left(3 + 2\,\lambda_4\,L^2\,g_+^\star \right) + 20\,\lambda_4\,L^2 \Big]\,.
\label{M_-}
\end{eqnarray}
where $\mathcal{M}_\pm^\star \equiv \mathcal{M}_\pm(g_\mp^\star,\,a_\star,\,\lambda_4,\,L^2)$ and $g_\pm^\star \equiv g_\pm (a_\star)$\,.
In addition, the solutions of the $g_\pm^\star = g_\pm(a_\star)$ functions are obtained by substituting $\mathcal{M}_\pm^\star$ into the following equations
\begin{eqnarray}
\Upsilon[g_\pm^\star] &=& -\frac{1}{L^2} + g_+^\star + \lambda_4\,L^2\,(g_+^\star)^2 = \frac{\mathcal{M}_\pm^\star}{a_\star^{3}}\,.
\end{eqnarray}
The functions $g_\pm^\star$ are given by
\begin{eqnarray}
g_\pm^\star &=& -\,\frac{(1 + \mathcal{C}_1) \pm \sqrt{1 + 4\,\lambda - 2\,\mathcal{C}_1 - 3\,\mathcal{C}_1^2 + 4\,\mathcal{C}_2\,\lambda\, L^2}}{2\,\lambda\,L^2}\,,
\label{g_pm}
\end{eqnarray}
where the coefficients $\mathcal{C}_{1,2}$ are defined by
\begin{eqnarray}
\mathcal{C}_1 = \frac{3\,a_\star^2\,(1+ 4\,\lambda_4)}{6\, a_\star^2 - 4\, \lambda_4\, L^2 }\,,
\qquad \quad
\mathcal{C}_2 = \frac{(1 + 4\,\lambda_4)\left(9 \,a_\star^2 + 20\,\lambda_4\, L^2 \right)}{4\,\lambda_4\, L^2 \left(3\,a_\star^2 - 2\, \lambda_4\, L^2 \right)}\,.
\end{eqnarray}
In the absent of the 4D EGB gravity by setting $\lambda_4\rightarrow 0$, we find that $g_-^\star$ has a good behavior (stable) while $g_+^\star$ gives infinite value (unstable). To study the phase transition between the exterior AdS ($\mathbb{M}_+$) manifold and the interior dS ($\mathbb{M}_-$), the condition $g_+^\star \neq g_-^\star$ is required.

It is worth summarizing about the junction conditions that we have so far. There are two constraints from the configurations of the thermalon as $V(a_\star)=0=V'(a_\star)$, Hawking temperature condition at the black hole radius ( $r_{BH}$ event horizon), $T = f'(r_{BH})/4\,\pi$ and the matching condition for the temperature of the thermal circle at the thermalon static configuration $\beta_+\sqrt{f_+(a_\star)}=\beta_-\sqrt{f_-(a_\star)}$. All of these conditions, there can be only one free parameter and in the following we choose $T_+=1/\beta_+$\,. In addition, these junctions will be useful when calculating of the Euclidean action is taken into account in the next section.   

Substituting the solutions of $\mathcal{M}_\pm^\star$ in Eqs.(\ref{M_+},\ref{M_-}) and the $g_\pm^\star$ in Eq.(\ref{g_pm}) to the effective potential in Eq.(\ref{V-eff}), we are ready to study the dynamics and the stability of the thermalon effective potential. The plots of the thermalon effective potential are depicted in Figure \ref{thermalon-potential} for $d=4$ and $d=5$ cases. We observe, in the plots, that in the $d\to 4$ limit, the effective potential is shallower than that of the $d=5$ theory as well as the existences of the potential is shorten. If the $\lambda_4$ coupling is getting bigger then the existences of the potential is getting closer to the thermalon position, see figure \ref{thermalon-potential}. In particular, the effective potential exists when the range of the $\lambda_4$ coupling is constrained as
\begin{eqnarray}
\lambda_4 \leq \frac{1}{8}\Big[ \left(4 a_\star^2-3 a_\star^4\right) + \sqrt{9 a_\star^8-24 a_\star^6+52 a_\star^4}\,\Big].
\end{eqnarray}
In this study, we set $a_\star =1$\,.  Then the upper bound of the 4D EGB coupling is $\lambda_4 \leq 0.885$\,. This result occurs only in a $4-$dimensional consideration in contrast to the $5$-dimensional EGB theory. For instance $a_\star = 1$ case, the effective potential exists for all positive values of the EGB coupling in $5$-dimensions.

We close this section by analysing the stability of the thermalon. The stability analysis of the thermalon is carried out in Refs.\cite{Camanho:2013uda,Camanho:2015ysa} (detail analysis and derivation are given) and numerically confirmed by Ref.\cite{Hennigar:2015mco} in a general $d-$dimensional EGB gravity and the thermalon is always expanding for all positive EGB coupling giving the phase transition. Here we follow their analysis in the $d\to 4$ limit and we do not repeat the results in the present work. The thermalon in the 4D EGB theory is always unstable under the perturbation around the thermalon configuration for all $\lambda_4>0$ and this leads to the thermalon escaping to infinity at the finite time as well as the $5$-dimension case. Here we demonstrate for $a_\star =1$ case in figure \ref{thermalon-potential} that there is no local minimum at the thermalon location. The thermalon is on the top of the effective potential and the phase transition is ready to take place for all existent 4D EGB coupling, $0 \leq \lambda_4 \leq 0.885$ for $a_\star =1$.

%%%%%%%%%%%%%%%%%%%%%%%%%%%%%%%%%%%%%%%
\section{AdS to dS phase transition in novel 4D EGB gravity}
\label{sec3}
%%%%%%%%%%%%%%%%%%%%%%%%%%%%%%%%%%%%%%%%
This section is aimed to examine the possibility of the thermalon mediated phase transition from asymptotically AdS to dS asymptotics by computing the difference between those two spacetimes. It is necessary to consider the event horizon and related physical quatities of the dS black hole in the interior spacetime which will be useful for computing the thermalon free energy in the latter. The thermalon (Euclidean sector of the thin-shell bubble) hosts the dS black hole in its interior with the exterior AdS asymptotics. This means that the location of the thermalon should be found between cosmological and event horizons of the interior dS space. To demonstrate this situation, we start with the evaluation of the black hole radius (event horizon). We begin with solving for $f_-(r_{H}) = 0$ in the function of the dS black hole mass inside the interior ($M_-$) as,
%%%%%%%%%%%%%%%%%%%%%%%%%%%%%%%%%%%%%%%%%%%%%%%%%%%%%%%%%%%%%%%%%%%%%%%%%%%%%
\begin{figure}[h]	
	\includegraphics[width=12.5cm]{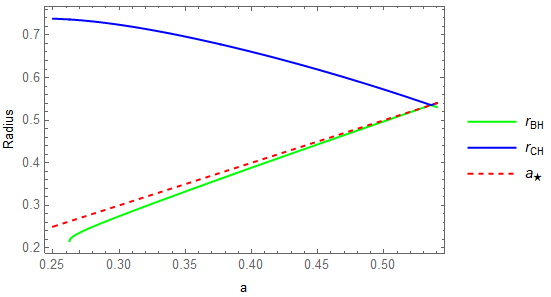}
	\centering
	\caption{The plot of the dS black hole radius, $r_{BH}$ (green), the cosmological horizon $r_{CH}$ (blue) as functions of $a_{\star}$ and the dashed red line is represented the thermalon (bubble) radius for $\lambda = 0.04$ and $L=1$. As shown in the plot, the thermalon location, $a_{\star}$, is always laid between the black hole radius, $r_{BH}$ and the cosmological horizon $ r_{CH}$. The meeting point of three lines is a so called the Nariai bound $a_{\rm Nariai}=\sqrt{1/\Lambda}=L/\sqrt{3}$ = $0.577$ for $L=1$.}
	\label{radius}
\end{figure}
%%%%%%%%%%%%%%%%%%%%%%%%%%%%%%%%%%%%%%%%%%%%%%%%%%%%%%%%%%%%%%%%%%%%%%%%%%%%
\begin{eqnarray}
f_-(r_{H}) = 0\,,\;\Rightarrow \; g_-(r_{H}) = \frac{1}{r_{BH}^2} \,,
\end{eqnarray}
where $r_H$ is the radius of the existent horizons of the spacetime. This leads to 
\begin{eqnarray}
r_H^4 - L^2\, r_H^2 + L^2\,\mathcal{M}_-\,r_H - L^4\,\lambda_4  = 0\,,
\label{quartic}
\end{eqnarray}
where the expression of the $\mathcal{M}_- \equiv \mathcal{M}_-(g_+(a),\,a,\,\lambda,\,L^2)$ is given by Eq.(\ref{M_-}). The above equation is written as a depressed quartic equation. There are existences of the two real positive roots and we can identify the roots as the radius of the black hole ($r_{BH}$) and cosmological horizon ($r_{CH}$). They are given by
\begin{eqnarray}
r_{BH} &=&\frac{1}{2\,\sqrt{6}}\left( \delta^2 -\sqrt{12 - \delta - 12 \sqrt{6}\,\frac{\mathcal{M}_-}{\delta}} \,\right),
\label{event-hor}
\\
r_{CH} &=&\frac{1}{2\,\sqrt{6}}\left( \delta^2 +\sqrt{12 - \delta - 12 \sqrt{6}\,\frac{\mathcal{M}_-}{\delta}} \,\right),
\label{cosmo-hor}
\\
\delta &=&  4 + 2^{2/3} \Delta + \frac{2 \,2^{1/3}}{\Delta}\, (1-12\,\lambda_4 )\,,
\\
\Delta &=& \left(27\,\mathcal{M}_-^2 - 2 - 72\,\lambda_4 -\sqrt{4\, (12\,\lambda_4 -1)^3 + \Big( 2 + 72\,\lambda_4 -27\,\mathcal{M}_-^2\Big)^2} \right)^{\frac13}\,.
\end{eqnarray}
We note that the black hole radius and cosmological horizon are written in a very complicated form. The plot of $r_{BH}$, $r_{CH}$ and $a$ (thermalon location) as functions of $a$ is displayed in figure \ref{radius}. According to the plot, the results confirm that the thermalon covers the black hole radius and the cosmological radius hosts the thermalon in the interior dS manifold. Due to the complicated expression of the black hole radius as shown in Eq.(\ref{event-hor}), this requires the range of the $\lambda_4$ parameter which gives the black hole radius becoming physical (real positive) values. We can not provide analytical expression of parameter range but we do it numerically instead. For $L=1$ and $a_\star =1$, it is found that $0 < \lambda_4 < 0.095$\,.

According to the existing literature, the thermalon mediated phase transition is computed in the spirit of the Hawking-Page (HP) procedure. We will calculate the free energies of thermalon hosting dS black hole in the interior space and the exterior thermal AdS asymptotics by evaluating the total Euclidean action of the system. In general, the Euclidean action naively gives the divergence due to the infinite volume of the entire space integration. But one can regulate the Euclidean action by using the background subtraction scheme. The final result allows the computation of the (semi-classical) Euclidean action equal to the product of the inverse Hawking temperature and free energy. However, in contrast to the HP phase transition, we use the free energy of the thermalon instead of the black hole while the subtraction background is the same as the HP transition, i.e., the AdS asymptotics. It has been demonstrated and calculated by Refs. \cite{Camanho:2012da,Camanho:2013uda,Camanho:2015ysa} that the total Euclidean action of the thermalon mediated AdS to dS phase transition in the higher order gravity is composed of the actions from the interior, the exterior manifolds and the bubble thin-shell. Having used the on-shell method with the background subtraction and imposed the junction conditions in the previous section, the Euclidean action reduces to a simple form,
\begin{eqnarray}
\mathcal{I}_E = \beta_+\,\mathcal{M}_+ - S\,, 
\label{Euclid-action}
\end{eqnarray}
where $\beta_-$ and $\mathcal{M}_-$ are the inverse Hawking temperature and the mass parameter of the exteriror observer at the AdS asymptotics while $S$ is entropy of the dS black hole. The result of the above equation can be rewritten in terms of the free energy equation in the canonical ensemble as
\begin{eqnarray}
\mathcal{F} = \mathcal{M}_+ - T_-\,S\,,
\label{free-energy}
\end{eqnarray}
where $\mathcal{M}_+ \equiv \mathcal{M}_+\big(g_-(a),\,a,\,\lambda_4,\,L^2\big)$ is equivalent to the enthalpy of the system. According to the matching condition of the thermal circle, the Hawking temperature of the exterior, $T_+$ reads
\begin{eqnarray}
T_+ = \sqrt{\frac{f_+(a_\star)}{f_-(a_\star)}}\,T_-\,,
\label{T_+}
\end{eqnarray}
where the $T_-$ is the Hawking temperature of the dS black hole hosted by thermalon in the interior and it is written by
\begin{eqnarray}
T_- = \frac{1}{4\,\pi}\frac{d\,f_-}{d\,r}\,\bigg|_{r= r_{BH}}\,,
\label{T_-}
\end{eqnarray}
where the explicit form of the metric function $f_-=f_-(r)$ is given by Eq.(\ref{metric-f-pm}). The entropy of the dS black hole can be calculated using a standard method and we find at $d\to 4$ limit \cite{Camanho:2013uda,Camanho:2015ysa,Hennigar:2015mco}\,,
\begin{eqnarray}
S = 4\,\pi\,\left(\frac{1}{2}\,r_{BH}^2 + 2\,\lambda _4\right).
\label{entropy}
\end{eqnarray}
%%%%%%%%%%%%%%%%%%%%%%%%%%%%%%%%%%%%%%%%%%%%%%%%%%%%%%%%%%%%%%%%%%%%%%%%%%%%%%%
\begin{figure}[!h]	
	\includegraphics[width=10cm]{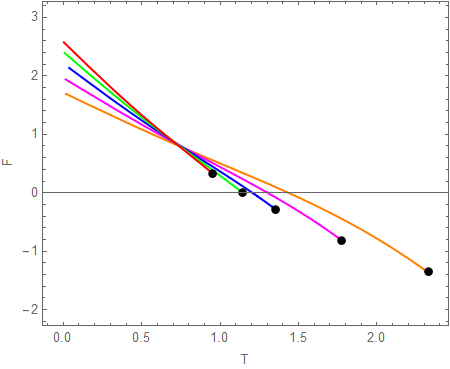}
	\centering
	\caption{The plot shows free energy $F$ associated to the thermalon configuration as a function of the temperature $T=\beta^{-1}_{+}$ by varying the values of the coupling $\lambda$ for $L=1$. From right to left: $\lambda_4a=0.04$ (orange), $\lambda_4=0.005$ (pink), $\lambda_4=0.06$ (blue), $\lambda_4=0.0682$ (green) and $\lambda_4=0.075$ (red). The black dots at the end of each plots represent the maximum temperature of the thermalon. All curves correspond to a physical solutions $\Pi^+ = \Pi^-$ due to the junction condition of thermalon in Eqs.( \ref{junction-EGB}) and (\ref{junction-Veff}).}
	\label{FT-plot}
\end{figure}
%%%%%%%%%%%%%%%%%%%%%%%%%%%%%%%%%%%%%%%%%%%%%%%%%%%%%%%%%%%%%%%%%%%%%%%%%%%%%%%
Substituting all related quantities from Eqs.(\ref{M_+}), (\ref{T_+}) and (\ref{entropy}) into the free energy equation Eq.(\ref{free-energy}), we are ready to study the phase transition structure in the phase space diagram. In addition, we choose the thermalon radius, $a$ as parameter in the phase space. Therefore, all physical quantities are related to the dS black hole radius should express in terms of $a$ via the relation in Eq.(\ref{event-hor}). 

In this work, we employ the technical calculation from Refs.
\cite{Camanho:2012da,Camanho:2013uda,Camanho:2015zqa,Hennigar:2015mco} that the free energy corresponded to the thermalon hosting the dS black hole in its interior is computed and then compare it to the thermal AdS of the exterior at the same temperature. More importantly, as done in the literature of the thermalon mediated phase transition, the free energy of thermal AdS space is set to zero ($F_{\rm AdS} = 0$) since it is used as the background subtraction for the Euclidean action. This indicates that the phase transition takes place when the free energy become negative at some values of the temperature, while if the free energy is positive the system is in a metastable state.  

The plot of the free energy ($F$) against temperture $(T_-)$ is depicted in figure \ref{FT-plot}. When setting $L=1$, all curves from right to left correspond to  $\lambda_4=0.04$ (orange), $\lambda_4=0.005$ (pink), $\lambda_4=0.06$ (blue), $\lambda_4=0.0682$ (green) and $\lambda_4=0.075$ (red) where the black dots at the end of each curve represent the maximum temperature of the thermalon. In addition, it is important to note that all curves correspond to a physical branch solutions for $\Pi^+ = \Pi^-$ due to the junction condition of thermalon in Eqs.( \ref{junction-EGB}) and (\ref{junction-Veff}) while the unphyiscal solution for $\Pi^+ = -\Pi^-$ are absent in the plot. As a result, the free energy for $\lambda_4 = 0.0682$ is equal to zero with the maximum temperature. This implies the existence of the phase transition taking place when $0 < \lambda_4 < 0.0682$. In addition, we also further investigate the upper bound of the $\lambda_4$ which makes the temperature becoming physical, i.e., $T_- > 0$ and discover $0 < \lambda_4 < 0.0834$\,. According to our results, we conclude that the profile of the thermalon mediated phase transition from AdS to dS asymptotics in the novel 4D EGB gravity is similar to that of the EGB gravity in $5$-dimension. In this scenario, there are two different asymptotic branch solutions as AdS and dS. At some particular values of the 4D EGB coupling, the thermal AdS is initially in the false vacuum state or metastable state and then decays into black hole inside dS space (true vacuum) through the nucleation of the thermalon. Eventually, when the phase transition occurs the thermalon of the interior dS space will infinitely expand and occupy a whole space changing the boundaries from AdS to dS. In another word, the phase transition can chang the effective cosmological constant from the negative to positive via the thermalon mediation. 

\section{Conclusion  \label{summary}}

In this work, we studied the possible existence of thermal phase transition between AdS to dS asymptotic geometries in vacuum in the context of novel 4D Einstein-Gauss-Bonnet (EGB) gravity. The phase transition proceeded through the thermalon (the Euclidean sector of the bubble thin-shell) formation having a black hole inside the interior of dS spacetime and a thermal AdS spacetime in the exterior without invoking any matter fields. According to our analysis, we discovered that the gravitational phase transitions in novel 4D EGB gravity took place for all existences of the 4D EGB coupling with $0 < \lambda_4 < 0.0682$. In contrast to 5-dimensional EGB gravity, the existence of phase transition required a narrow range of the 4D EGB coupling with particular critical values, while the 5-dimensional sector has a wider range of its coupling values for the existences of the thermalon and there also exist critical values for the emergence of the phase transition. However, the profile of the thermalon mediated phase transition from AdS to dS asymptotics in the novel 4D EGB gravity is similar to that of the EGB gravity in $5$-dimensions.

At some particular values of the 4D EGB coupling, the thermal AdS is initially in the false vacuum state or metastable state and then decays into black hole inside dS space (true vacuum) through the nucleation of the thermalon. Eventually, when the phase transition occurs the thermalon of the interior dS space will infinitely expand and occupy a whole space changing the boundaries from AdS to dS. To be more precise, the phase transition can change the effective cosmological constant from the negative values to positive ones via the thermalon mediation. 

The difference between the phase transition in $4$-dimensional and $5$-dimensional EGB theory is worth mentioning. Although, the phenomenon of the phase transition dose not change in this model when we change from $5$ to $4$ dimensional theories. However, the viable range of the higher order gravitational couplings resulting the existence of the phase transition are quite different. The coupling of the novel 4D EGB gravity is narrower and smaller compared with the traditional EGB gravity in $5$ dimensions in which the EGB coupling has a wider range. This behavior may be speculated that the rescaling affects the EGB coupling in $4$ dimensions. Further investigations on the dimensionality of the EGB gravity in another physical phenomena both quantum and classical levels may be fruitful and useful for a deeper understanding of the Gauss-Bonnet gravity in lower dimension.   

Interestingly, thermal phase transition between AdS to dS asymptotic geometries in vacuum in the context of novel 4D Einstein-Gauss-Bonnet (EGB) gravity in the present of charged sectors in also worth investigating, see Refs.\cite{Samart:2020qya,Samart:2020mnn}. However, we leave this interesting issue for further investigation.

\end{document}